\documentclass[%
 reprint,
 amsmath,amssymb,
 aps,
 pra,
]{revtex4-2}

\usepackage{graphicx}
\usepackage{dcolumn}
\usepackage{bm}
\usepackage{physics}
\usepackage{svg}
\usepackage{multirow}
\usepackage{xcolor}
\usepackage{makecell}
\usepackage{longtable}
\usepackage{booktabs}
\usepackage{float}

\newcommand{\eps}{\epsilon}

\newcommand{\veps}{\varepsilon}
\newcommand{\ox}{\otimes}

\begin{document}

\preprint{APS/123-QED}

\title{Smolyak algorithm assisted robust control for quantum systems with uncertainties}

\author{Zigui Zhang}
\author{Zibo Miao*}%
\affiliation{%
School of Intelligence Science and Engineering, Harbin Institute of Technology, Shenzhen, 518055, China
}%


\author{Xiu-Hao Deng}
\affiliation{Shenzhen Institute of Quantum Science and Engineering, Southern University of Science and Technology, Shenzhen, 518055, China}
\affiliation{International Quantum Academy, Shenzhen, 518000, China}

\date{\today}

\begin{abstract}
Efficient and systematic numerical methods for robust control design are crucial in quantum systems due to inevitable uncertainties or disturbances. We propose a novel approach that models uncertainties as random variables and quantifies robustness using the expectation of infidelity by reformulating it as a weighted tensor product quadrature. We employ the Smolyak algorithm to develop a parametric robust quantum control scheme, which balances the reduction of computational cost with the enhancement of estimation accuracy.
We demonstrate the effectiveness of our proposed algorithm by incorporating the Smolyak sparse grids into conventional gradient-based quantum optimal control methods such as GRAPE and GOAT. In robust control problems concerning quantum gate realization, low infidelity and strong robustness can be achieved. These results contribute to improving the reliability and security of quantum computing and communication systems in the presence of real-world imperfections.
\end{abstract}

\maketitle

\section{Introduction}\label{sec:introduction}
Quantum control is crucial to the successful realization of quantum technologies such as quantum computation, which is usually implemented by manipulating the Hamiltonians of quantum systems. In particular, developing a systematic control framework that can robustly drive the dynamics of the quantum system towards the target in the presence of modeled noise or unknown uncertainty remains a challenge \cite{kochQuanOptiCtrl2022}. Many efforts have been made to improve the robustness of quantum control in various ways. For example, the geometric correspondence of the evolving manifold, such as curvature, can be exploited to improve the robustness \cite{zengGeoForm2019,buterakosGeoForm2021,dongDoubleGeoQuanCtrl2021,barnesDynmCorrGeo2022,haiGCRC2022,yiCompoPulse2024}. Similarly, by elaborately constructing the dynamical trajectory of the system, the noise terms in the expansion of the objective function can be eliminated to a great extent\cite{daemsQuanRobustCtrl2013,dridiRIO2020,laforguePulseAnalytic2022}, or others \cite{robertSCP2013,vanRobustTwoLevel2017,poggiMinNorm2024}. In addition, methods from other perspectives have been proposed, such as dynamical decoupling \cite{lorenzaDD1999,GenkoDD2017}, composite pulses \cite{genovCompoPulse2014,wuCompoPulse2023} and shortcuts to adiabaticity \cite{guShortAdiabat2019}.

In most cases, the analytic form of control is rather complicated. For some expansion-based methods, it is troublesome to calculate high-order expansions to achieve high-level robustness. In this regard, an effective numerical approach without complex expansion to achieve high-level robustness may be a better solution. A promising candidate is the sampling-based (ensemble-based) method \cite{chens-GRAPE2014}. By modeling the uncertainty that inevitably exists in quantum systems as random variables following prescribed distributions, we can evaluate the robustness via expectation of the control objective as a performance index. 
To this end, the performance index can be approximated by averaging or performing a weighted sum of the samples, which are obtained by different strategies, such as dense grids \cite{chens-GRAPE2014,andersonNum2015,dongSLC2015,dongSLC2020}, Monte Carlo methods \cite{wub-GRAPE2019,geRisk2021}, or other algorithms \cite{geLearnA-GRAPE2020}.

However, the number of samples grows exponentially due to the curse of dimensionality, which leads to huge costs in computation. The Monte Carlo sampling strategy can reduce the sampling cost, but it tends to be slow in convergence and unstable for small samples.
Although an efficient control strategy can be appropriately designed by learning-based algorithms, it requires modification or retraining when the setting of dynamics changes or more uncertainties are involved. Therefore, a reliable and universal sampling strategy is still in need.

Inspired by sampling-based algorithms, in this paper we take advantage of the Smolyak sparse grids \cite{smolyak1963} to develop a generic robust quantum control scheme in the presence of uncertainties. Considering robustness as a weighted tensor product quadrature, higher accuracy and reduced computational cost can be achieved. 
In the remainder of the paper, we first introduce the mathematical model of quantum robust optimal control and provide different sparse grid assisted methods. The features and superiority of the proposed scheme are demonstrated in Sections \ref{sec:simulation} in the context of robust control problems. 
To be specific, by utilizing the Smolyak sparse grids in estimation and optimized by gradient-based quantum optimal control methods, the robustness of single-qubit gates (the Hadamard gate, the $\pi/8$ gate, the phase gate, and the $R_x(\pi)$ gate) or the CNOT gate can be enhanced, particularly for low-to-medium uncertainty scales. 
Finally, Section \ref{sec:conclusion} provides the concluding remarks.

\section{Quantum robust control based on the Smolyak algorithm}\label{sec:RQCBOTSA}

We first introduce the mathematical framework of quantum robust control.

\subsection{Quantum robust control as multivariate quadrature}

Generally, the goal of quantum robust control is to find a specific control field $u^*$ that can guide the dynamics toward the target state or the operator over a time interval $T$ in the presence of some uncertainties $\delta$. In particular, if uncertainties $\delta$ can be parameterized as a function $f$ of instant $t$ and some random variables $\veps$, i.e. $\delta=f(t,\veps)$, then the performance index $J$ of robust control can be expressed as
\begin{equation} 
    J(u) := \mathbb{E}[J_\delta(u)]=\int_{\Omega} P(\veps)J_{f(t,\veps)}(u){\rm d} \veps,\label{eq:paraeps}
\end{equation}
with a given distribution $P(\veps)$ and spaces $\Omega$ of random variables $\veps$. Then the optimal control field can be obtained by $u^*=\arg \min_{u} J(u)$ with some quantum optimal control algorithms.

Given fixed values of the uncertainties $\delta$, the control performance $J_\delta(u)$ can be quantified by evaluating the discrepancy between the target and the dynamics that evolve over a time interval $T$. For quantum systems with uncertainties involved, whose dynamics obey the following Schr\"{o}dinger equation:
\begin{equation} 
    \dv{U(t)}{t}=-iH_{\rm{tot}}(t,u,\delta)U(t).\label{eq:schrodinger_eq}
\end{equation}
The total Hamiltonian $H_{\rm{tot}}$ is given by
\begin{equation} 
    H_{\rm{tot}}(t,u,\delta)=H_0+\sum_ju_j(t)H_{cj}+\sum_l\delta_l(t)V_l,
\end{equation}
where $H_0$, $H_{cj}$, $V_l$ represent the free Hamiltonian, the Hamiltonian corresponding to the control signal $u_j(t)$, and the Hamiltonian related to the uncertainty $\delta_l$, respectively. 
Let the target propagator be $U_{\rm{F}}$, the Frobenius operator infidelity is defined as
\begin{equation}  
    \Phi_1(U_{\rm F},U(T))=||U_{\rm F}-U(T)||^2,\label{eq:infide_norm2}
\end{equation}
with $||U||^2:=\Tr(U^\dag U)$ for any unitary operator $U$. In fact, an arbitrary global phase $\varphi$ is allowed for the system to approach the desired operator $U_{\rm{F}}$, that is \cite{navinGRAPE2005}
\begin{equation}  
    \Phi_2(U_{\rm F},U(T))=\min_\varphi||U_{\rm F}-e^{i\varphi}U(T)||^2.\label{eq:infide_normphase2}
\end{equation}
Note that the fidelity provided by Eq.~\eqref{eq:infide_normphase2} is not normalized, and a commonly used alternate is given by
\begin{equation}  
    \Phi_3(U_{\rm F},U(T))=1-\left|\frac1{\rm{dim}}\Tr(U_{\rm F}^\dag U(T))\right|^2,\label{eq:infide_corr2}
\end{equation}
where $\dim$ denotes the dimension of the system. 

In this context, the scenario considered here is analogous to that of an open quantum system, where dynamics are typically described by the Lindblad master equation and infidelity is evaluated by some functions of the density operator $\rho$.

In most cases, it is too time consuming or complicated to obtain an explicit analytic form of the performance index Eq.~\eqref{eq:paraeps}, and thus it is reasonable to consider a subset of $\Omega$ as an approximation. 
Mathematically, ${\mathbb E}[J_\delta(u)]$ can be viewed as a multidimensional integration with the weighting function $P(\veps)$, thus it can be approximated by a multivariate quadrature of the weighted tensor product
\begin{eqnarray} 
    J(u)&&\approx(Q_1^{n_1}\ox \cdots\ox Q_d^{n_d})J_\delta(u)\nonumber\\
    &&=\sum_{j_1=1}^{n_1}\cdots\sum_{j_d=1}^{n_d} w_{1,j_1}\cdots w_{d,j_d}J_{f(t,\veps_{1,j_1},\cdots,\veps_{d,j_d})}(u),
\end{eqnarray}
where $\ox$ denotes the tensor product, and the one-dimensional quadrature $Q_k^{n_k}$ with order $n_k$ of the $k$-th variable provides a set of nodes $\{\veps_{k,j_k}\}$ and the corresponding weights $w_{k,j_k}$ $(1\le j_k\le n_k)$. A different choice of the quadrature rule gives rise to a different sampling set $\mathcal{S}=\{\veps=(\veps_{1,j_1},\cdots,\veps_{d,j_d})|1\le j_k\le n_k,1\le k\le d\}$. 
In addition to quadrature-based methods, other approximation techniques utilizing sampling sets are acceptable. Such approaches can collectively be referred to as sampling-based algorithms \cite{chens-GRAPE2014,wub-GRAPE2019}.

\subsection{Smolyak algorithm assisted quantum robust control}\label{sec:smROC}

Smolyak algorithm provides an efficient approach to reducing the number of nodes in the tensor product. We will show how it can be applied to quantum robust control in this section. To be specific, for a linear operator $L$ with a convergent approximation sequence $L^j(j=0,1,\cdots)$, the tensor product of such operators can be written in terms of difference operators, namely
\begin{eqnarray} 
    && L_1\ox \cdots\ox L_d=\sum_{j_1=0}^{\infty} \Delta_1^{j_1}\ox \cdots \ox\sum_{j_d=0}^{\infty}\Delta_d^{j_d},\\ 
    &&\textrm{where}\quad 
    \begin{cases}
        \Delta^{j_k}_k=L^{j_k}_k-L^{j_k-1}_k&(j_k\ge 1)\\
        \Delta^{0}_k=L^0_k=0&(j_k=0)
    \end{cases}\nonumber
\end{eqnarray}

The essence of the Smolyak algorithm can be interpreted as finding a proper subset $\mathcal{I}$ of multi-index $(j_1,\cdots,j_d)$, called the Smolyak sparse grids, to approximate the tensor product as precisely as possible while reducing the computational cost. Particularly, the Smolyak-type quadrature of the following form,
\begin{equation} 
    A(q,d)=\sum_{(j_1,\cdots,j_d)\in \mathcal {I}(q,d)}\Delta_1^{j_1}\ox \cdots\ox\Delta_d^{j_d},
    \label{eq:Aqd}
\end{equation}
can provide a subset $\mathcal{I}$ and the corresponding upper error bound when the order of the sequence $L^j$ is at most $q$, with $\mathcal{I}(q,d)=\{(j_1,\cdots,j_d):j_1+\cdots+j_d \le q\}$. It is more convenient in programming to reformulate Eq.~\eqref{eq:Aqd} as \cite{wasilkowskiBoundTP1995}:
\begin{eqnarray} 
    A(K,d)=&&\sum_{\max(K,d)\le S_j\le K+d-1} (-1)^{K+d-1-S_j}\nonumber\\
    &&\binom{d-1}{S_j-K}L_1^{j_1}\ox\cdots\ox L_d^{j_d}.\label{eq:smolyak_sparse_grid}
\end{eqnarray}
Here, $S_j=j_1+\cdots+j_d \;(j_k\ge 1)$, and $\binom{d-1}{S_j-K}$ is a binomial number. The parameter $K=q-(d-1)$ defines the sparse grid level. Generally, the precision of Eq.~\eqref{eq:smolyak_sparse_grid} can increase as the sparse level $K$ increases.

\begin{figure}[htbp!]
    \centering
    \includegraphics[width=0.45\textwidth]{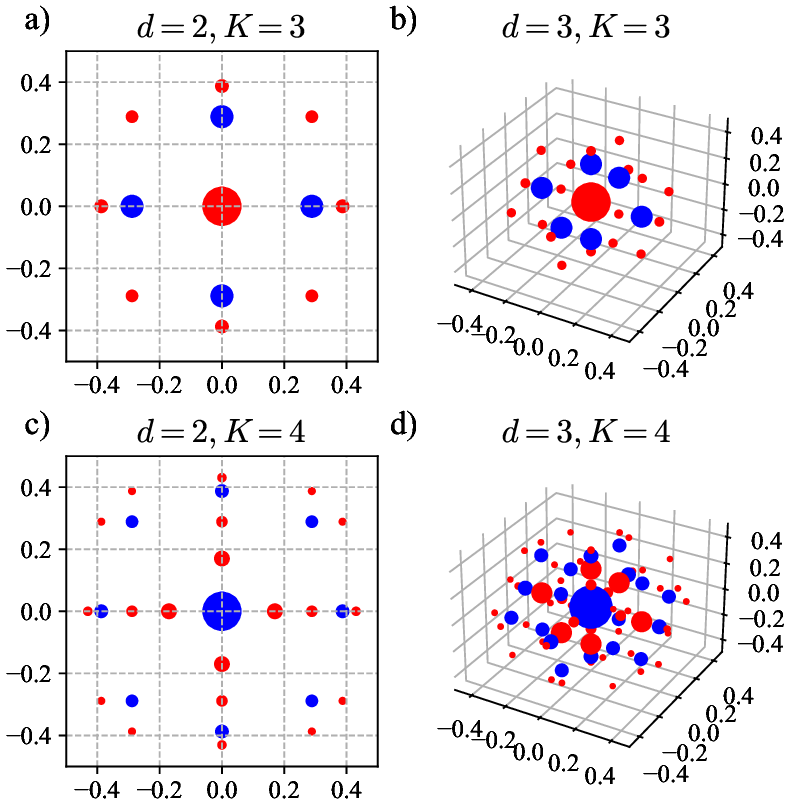}
    \caption{Demonstration of abscissas (nodes) and weights of the Smolyak sparse grids with respect to Gauss-Legendre quadrature rule, when (a) $d$=2 and $K$=3,  (b) $d$=3 and $K$=3, (c) $d$=2 and $K$=4, (d) $d$=3 and $K$=4, respectively (the number of random variables that take values according to the axes is $d$). The color and size of circles correspond to the signs of weights and their absolute values  (the color blue and red correspond to the minus and positive signs respectively).}
    \label{fig:smolyak_weighting_samples}
\end{figure}

Therefore, the performance index $J(u)$ defined in Eq.~\eqref{eq:paraeps} can then be approximated by
\begin{equation} 
    J(u)\approx A(K,d)J_\delta(u)=\sum_{\veps\in S(K,d)} w_{\veps}J_{f(t,\eps)}(u),
\end{equation}
where the abscissas (nodes) $\veps$ are taken from the Smolyak sparse grid $\mathcal{S}(K,d)=\{(\veps_{1,j_1},\cdots,\veps_{d,j_d}):(j_1,\cdots,j_d)\in\mathcal{I}(q,d)\}$, with the corresponding weight $w_{\veps}$. Especially, when $L$ is taken as the Gaussian quadrature, the quadrature $A(K,d)(J)$ is precise for all polynomials defined on the spaces $P_{k_1}\ox\cdots\ox P_{k_d}$ with $k_1+\cdots+k_d =2K-1$. 
We take the operator $L$ as the Gauss-Legendre quadrature over the interval $[-0.5,0.5]$ with the weighting function $P(\veps)\equiv 1$ for demonstration. In different settings $d,K$ as shown in Figure~\ref{fig:smolyak_weighting_samples}, the associated abscissas and the weights of the Smolyak sparse grids are distributed sparsely in space and more dense as $K$ increased.

The dense grids adopted in \cite{chens-GRAPE2014,dongSLC2015,dongSLC2020} require $O(n^d)$ points, with $n$ nodes in each dimension, making them computationally expensive due to exponential growth. To alleviate the computation burden, the Monte Carlo method takes the following approximation
\begin{equation} 
    J(u)\approx \frac1N\sum_{\varepsilon\in S_b} J_{f(\varepsilon)}(u),\label{eq:monte-carlo}
\end{equation}
where the sampling set $S_b$ comprises $N$ samples drawn from the probability distribution $P(\varepsilon)$. However, its static error is of the order $O(1/\sqrt{N})$, leading to instability in the final turn of optimization \cite{wub-GRAPE2019}. 
In the case of low-to-medium uncertainty scales ($d\le 10$), the Smolyak algorithm employs significantly fewer points compared to dense grids, namely $O\left(n(\log n)^{d-1}\right)$ points. Meanwhile, it provides a more accurate estimate than the Monte Carlo method, which mitigates the instability in optimization. It can be concluded that a balance between efficiency and accuracy can be achieved by adopting the Smolyak sparse grid $\mathcal{S}(K,d)$ as the sampling set.

However, the Smolyak algorithm demands higher smoothness of the integrand with respect to uncertainties, i.e., bounded mixed derivatives $\partial^\alpha{J_\delta}/\partial{\veps}^\alpha$ for orders $\alpha=(\alpha_1,\ldots,\alpha_d)$. In the absence of smoothness (e.g., non-smoothness observed in certain complex quantum systems), its computational advantages could be diminished to a great extent. This issue may be addressed by avoiding negative weights in sparse grid construction or employing adaptive sparse grids.

Additionally, the $n$-th order raw moment of the performance index, denoted by ${\mathbb E}[J^n_\delta(u)]$, can be estimated using the Smolyak sparse grid $A(K,d)(J_\delta^n)$. For example, a higher sparse level may be required to ensure precision while calculating variance $D[J_\delta(u)]={\mathbb E}[J^2_\delta(u)]-{\mathbb E}^2[J_\delta(u)]$. Detailed analyses of the computational error and cost of the Smolyak algorithm can be found in \cite{wasilkowskiBoundTP1995,wasilkowskiWTP1999}. 

The related algorithms to optimize the performance index can generally be divided into two classes. One is the class of gradient-based methods, such as widely used \textbf{GRAPE} \cite{navinGRAPE2005} and \textbf{GOAT} \cite{machnesGOAT2018}. The other is the class of gradient-free methods such as \textbf{CRAB} \cite{canevaCRAB2011}. We provide several possible applications of the Smolyak sparse grids, for instance the cost function in \textbf{CRAB} can be rewritten as
\begin{equation} 
    \mathcal {F}=\alpha A(K,d)(\Phi)+\sum_j\beta_j\mathcal{C}_j[u_j(t)],
\end{equation}
for some coefficients $\alpha$ and $\beta_j$ and constraints $\mathcal{C}_j$. In the implementation of gradient-based methods, one can take
\begin{equation} 
    \pdv{{\mathbb E}[J_\delta(u)]}{u(t)}\approx \pdv{A(K,d)J_\delta}{u(t)}=\sum_{\veps\in S(K,d)} w_{\veps}\pdv{J_{f(\eps)}(u)}{u(t)}\label{eq:sm_gradient}
\end{equation}
as an approximation of the gradient.


\begin{figure*}[thp!]
    \centering
    \includegraphics[width=0.75\textwidth]{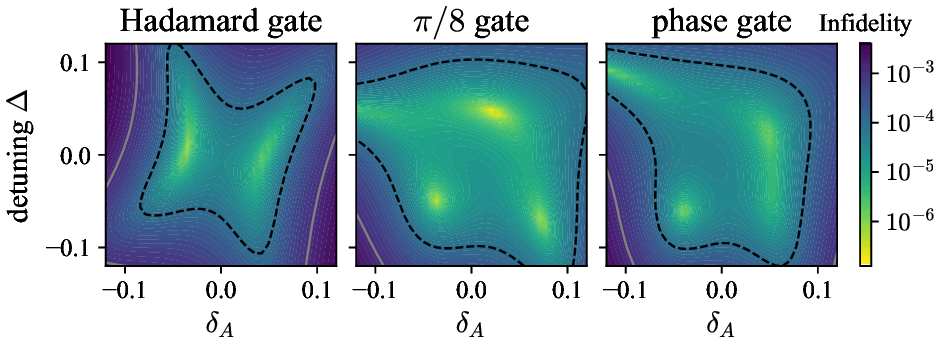}
    \caption{Landscapes quantifying the robustness, corresponding to the Hadamard gate $U_{\rm H}$, the $\pi/8$ gate $U_{\rm \pi/8}$, and the phase gate $U_{\rm S}$, respectively. The gray solid and black dashed lines represent the levels of robustness quantified by infidelity Eq.~\eqref{eq:infide_normphase2} at $10^{-3}$ and $10^{-4}$ respectively.}
    \label{fig:H_T_S_gate_pulse_and_robustness}
\end{figure*}

\section{Simulations}\label{sec:simulation}

\subsection{Robust realization of the Hadamard gate, $\pi/8$ gate and phase gate}
\label{subsec:Robust_single_qubit_gates}

Consider three gates of significant interest in quantum computing, namely the Hadamard gate $U_{\rm H}$ and the $\pi/8$ gate $U_{\pi/8}$ and the phase gate $U_{\rm S}$ \cite{nielsenQuanCompQuanInform2010}:
\begin{equation} 
    U_{\rm H}=\frac1{\sqrt2}\mqty[1&1\\1&-1],U_{\pi/8}=\mqty[1&0\\0&e^{i\pi/4}],U_{\rm S}=\mqty[1&0\\0&i].
\end{equation}
The system Hamiltonian steered by two control signals $u_x(t)$ and $u_y(t)$ can be given by:
\begin{equation} 
    H(t)=\frac{\Delta}2\sigma_z+(1+\delta_A)\left(\frac12u_x(t)\sigma_x+\frac12u_y(t)\sigma_y\right),
\end{equation}
where $\sigma_{x,y,z}$ are Pauli operators, $\Delta,\delta_A$ characterize detuning and inhomogeneousness of control pulse, which are quasi-static during the operation. Assuming that the uncertainties follow the same uniform distribution, i.e. $\Delta,\delta_A \sim \mathbb{U}(-0.1,0.1)$, we proceed with the Gauss-Legendre quadrature rule to evaluate performance index with infidelity defined in Eq.~\eqref{eq:infide_normphase2}. 

For practically convenience, the control signals are taken as smooth pulse in the form of Fourier series as follows:
\begin{align} 
    u_j(t)=G(t)&\left(a_{0,j}+\sum_{n_u=1}^N a_{n_u,j}\cos\left(\frac{2n\pi t}{T_p}\right)\right.\nonumber\\
    &\left.+\sum_{n_u=1}^Nb_{n_u,j}\sin\left(\frac{2n\pi t}{T_p}\right)  \right).
    \label{eq:Fourier_pulse_with_envelope}
\end{align}
The period time $T_p$ and the envelope function $G(t)$ should enforce the boundary conditions that $u_j(t)\rightarrow 0/\dv{u_j}{t}\rightarrow 0$ when $t\rightarrow 0/t\rightarrow T_p$. In particular, we use $G(t)=\sin^2\left(\frac{\pi t}{T_p}\right)$ as the pulse envelope to satisfy the above conditions. 

In the numerical simulation, we set $T=T_p=10$, $N=3$, and the sparse level $K=3$ that is adequate for the number of random variables $d=2$. Using the sampling set provided by the Smolyak sparse grid, we implement \text{GOAT} with L-BFGS-B \cite{lbfgsb1995} to optimize the control performance, referred to as \textbf{smGOAT} for short. The robust control landscape with respect to the Hadamard gate, the $\pi/8$ gate, and the phase gate can be observed in Figure~\ref{fig:H_T_S_gate_pulse_and_robustness}. The existing uncertainty, such as detuning or inhomogeneity of the control pulse, is detrimental to the control performance. With the help of \textbf{smGOAT}, the region enclosed by the robustness level $10^{-4}$ (black dashed line) is evidently expanded and the resulting average infidelity $\mathbb{E}[\Phi_2]$ for the three gates is around $1.87\times 10^{-4}$, $4.18\times 10^{-5}$, $7.35\times 10^{-5}$, respectively. This indeed verifies that \textbf{smGOAT} is able to achieve robust control protocols. The related parameters of the control signals are detailed in Table~\ref{tab:table1}.

\renewcommand{\arraystretch}{1.2} 
\begin{table*}[htbp!]
    \caption{\label{tab:table1}The Fourier coefficients of the control pulses for single-qubit gates.}

    \begin{ruledtabular}
    \begin{tabular}{cccll}

    Target & $K$ & Pulse & \{$a_j$\} & \{$b_j$\}\\

    \hline 

    \multirow{2}{*}{$U_{\rm H}$} & \multirow{2}{*}{3} & $u_x$ & [-1.10205484, - 0.16444018, 0.356119, 1.80099137] & [0.74186792, -1.1333456, -1.22726687]\\
    \cline{3-5}
    & & $u_y$ & [2.24002595, 3.0787707, -0.54292804, -1.32754733] & [2.01127864, 1.7822432, 1.51006954]\\

    \hline 

    \multirow{2}{*}{$U_{\pi/8}$} & \multirow{2}{*}{3} & $u_x$ & [-1.97064098, -4.01920656, 0.31203617, 0.8928809] & [1.06616553, 1.28528487, 0.20736214]\\\cline{3-5}
    & & $u_y$ & [2.39685112, 4.87430845, -0.56553063, -1.26091625] & [0.97969155, 1.14522837, 0.38519095]\\

    \hline 

    \multirow{2}{*}{$U_{\rm S}$} & \multirow{2}{*}{3} & $u_x$ & [-1.59137647, -3.52012761, 1.21916556, 1.10448505] & [-1.87008537, -2.18823133, -0.63684599]\\\cline{3-5}
    & & $u_y$ & [-1.28230799, -2.89252763, 0.8221858, 0.63479593] & [2.00875836, 2.18907434, 0.05213139]\\

    \hline 

    \multirow{4}{*}{$R_x(\pi)$} & \multirow{2}{*}{3} & $u_x$ & [0.5286695, 0.88444058, 1.27386703, 0.60613856] & [0.70152979, 1.12258258, 0.77939379]\\\cline{3-5}
    & & $u_y$ & [-0.9268824, -0.87491169, 0.15829603, 0.11848552] & [-0.45997346, -0.70470614, -0.35510744]\\
    \cline{2-5}
    & \multirow{2}{*}{4} & $u_x$ & [-0.46720312, -0.51100983, -1.41384471, -0.55770797] & [0.39306135, 0.7920471, 0.56493162]\\\cline{3-5}
    & & $u_y$ & [-1.329766, - 1.79146926, - 0.52194291, - 0.37675494] & [0.50968566, 0.77403372, 0.65863078]\\

    \end{tabular}
    \end{ruledtabular}
\end{table*}

\subsection{Robust realization of the $R_x(\pi)$ gate}

In terms of achieving robust implementation of the $R_x(\pi)$ gate, we first consider the scenario in which disturbances are allowed in all directions of the Pauli operators ($x$, $y$, and $z$). The system Hamiltonian can thus be written as
\begin{align} 
H(t)=&\frac12u_x(t)\sigma_x+\frac12u_y(t)\sigma_y\nonumber\\
&+\frac12\Delta_x\sigma_x+\frac12\Delta_y\sigma_y+\frac12\Delta_z\sigma_z,\label{eq:3_dir}
\end{align}
where $\Delta_{\{x,y,z\}}$ are independent disturbances, and the dynamics can be manipulated through $x$ and $y$. In this case, the $R_x(\pi)$ gate ($\pi$-pulse) rotates the qubit along the $x$ direction with an angle of $\pi$ and is supposed to be realized within the time interval $T=50$. Similarly, assuming $\Delta_{\{x,y,z\}}\sim \mathbb{U}(-0.05,0.05)$, we optimize the average infidelity defined in Eq.~(\ref{eq:infide_norm2}) with \textbf{smGOAT}. The control pulses are analogous to the previous example, where $N=3$ and the sparse level is taken as $K=3$ or $K=4$. 

\begin{figure*}[htbp!]
    \centering
    \includegraphics[width=0.8\textwidth]{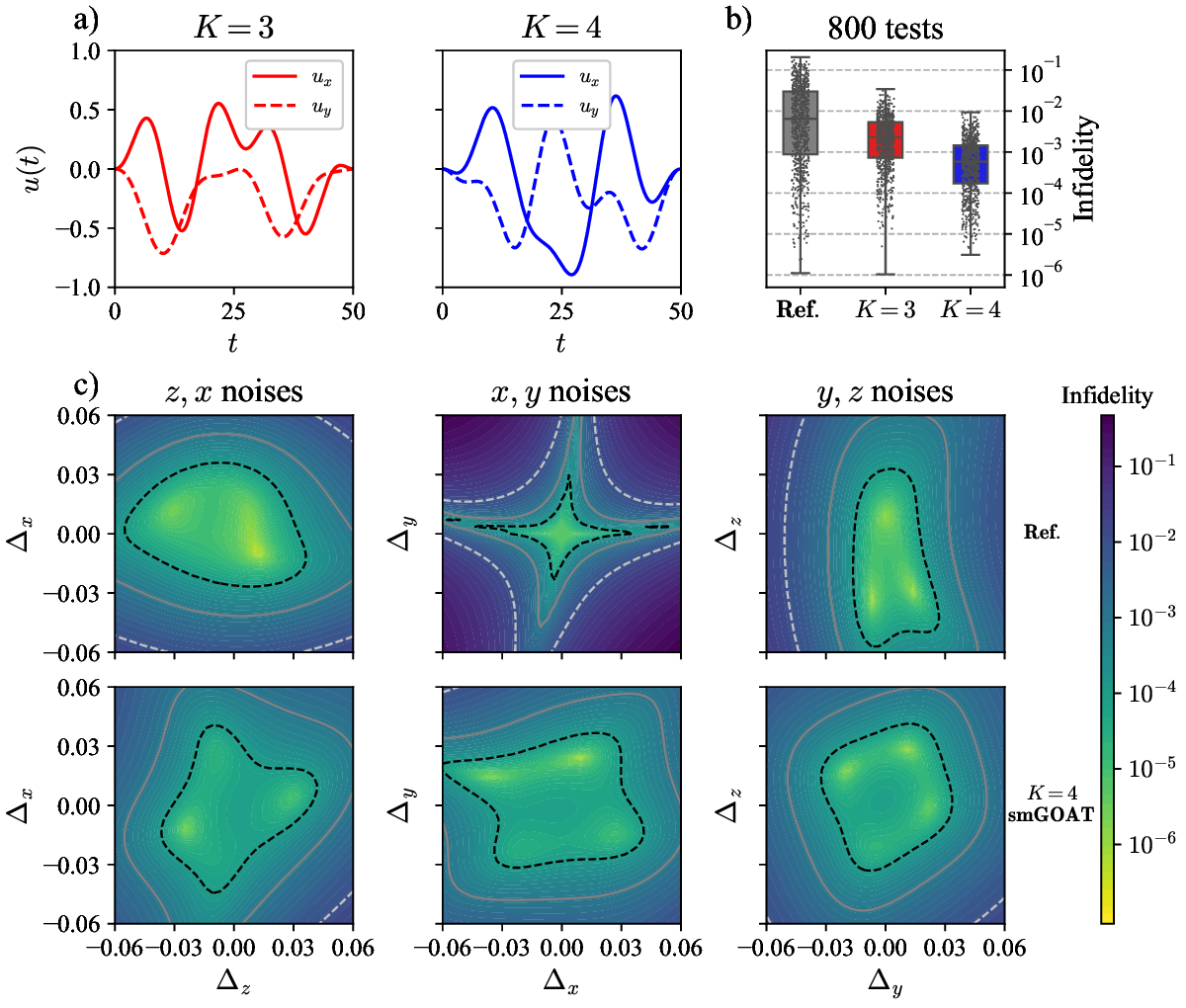}
     \caption{Simulation results regarding robust $R_x(\pi)$ gate. The resulting control pulses are plotted in (a).
     The level of robustness quantified by infidelity when noises or disturbances exist in two directions ($z-x$, $x-y$ and $y-z$) is plotted in (c), where the white dashed, gray solid and black dashed lines correspond to the infidelity values at $10^{-2}$, $10^{-3}$ and $10^{-4}$, respectively. 
     Infidelity through 800 tests of all directional disturbances uniformly sampled from cubic space $[-0.05,0.05]^3$ are box-plotted in (b).}
    \label{fig:pi_pulse}
\end{figure*}

The simulation results of robust $\pi$-pulse using \textbf{smGOAT} have been demonstrated in Figure~\ref{fig:pi_pulse}, compared to the geometric method (denoted by \textbf{Ref.}) discussed in \cite{haiGCRC2022,yiCompoPulse2024}.
To be specific, 800 tests of all disturbances are represented by the black dots in Figure~\ref{fig:pi_pulse} (b). It can be seen that the worst infidelity given by \textbf{smGOAT} is much better than that given by \textbf{Ref.}, which have been reduced 8 times for $K=3$ and 20 times for $K=4$, respectively. In addition, the average infidelity $\mathbb{E}[\Phi_1]$ provided by \textbf{smGOAT} is around $2.05\times 10^{-3}$ for $K=3$ and $6.5\times 10^{-4}$ for $K=4$. Furthermore, Figure~\ref{fig:pi_pulse} (c) shows the infidelity landscapes compared between the \textbf{Ref.} and \textbf{smGOAT} with $K=4$ in the presence of two directional disturbances. Obviously in the $x,y$ directions, the region enclosed by the robustness level $10^{-4}$ obtained by \textbf{smGOAT} is obviously larger.
Since the geometric method typically neglect higher-order mixed derivatives regarding the objective function, which leads to a squeezed and shrunken robust region as indicated in Figure~\ref{fig:pi_pulse} (c) when there are noises in the $x$ and $y$ directions. This example indicates the advance of the sampling-based method in obtaining stronger robustness when multiple disturbances are involved.

To demonstrate other applications of Smolyak sparse grids, the gradient-based \textbf{GRAPE} with the Smolyak algorithm (we use Eq.~\eqref{eq:sm_gradient} to calculate the gradient), called \textbf{smGRAPE} for short, will be used to optimize the control performance. It is worth noting that \textbf{GRAPE} can be accelerated by Adam \cite{kingmaAdam2014}, which is thus integrated into \textbf{smGRAPE}. In order to illustrate the ability of our approach to handle complicated uncertainties, we introduce two additional inhomogeneities, $\delta_x,\delta_y$, into the control fields, namely
\begin{eqnarray}
    H(t)=&\frac12(1+\delta_x)u_x(t)\sigma_x+\frac12(1+\delta_y)u_y(t)\sigma_y\nonumber\\
    &+\frac12\Delta_x\sigma_x+\frac12\Delta_y\sigma_y+\frac12\Delta_z\sigma_z,
\end{eqnarray}
such that the number of uncertainties is $d=5$. Similarly, we set $\delta_{\{x,y\}},\Delta_{\{x,y,z\}}\sim \mathbb{U}(-0.05,0.05)$, $T=50$, adopt the performance index defined in Eq.~\eqref{eq:infide_corr2} and set the time duration segmentation to 100 for the piecewise control protocol.

\begin{figure}[htbp!]
    \centering
    \includegraphics[width=0.36\textwidth]{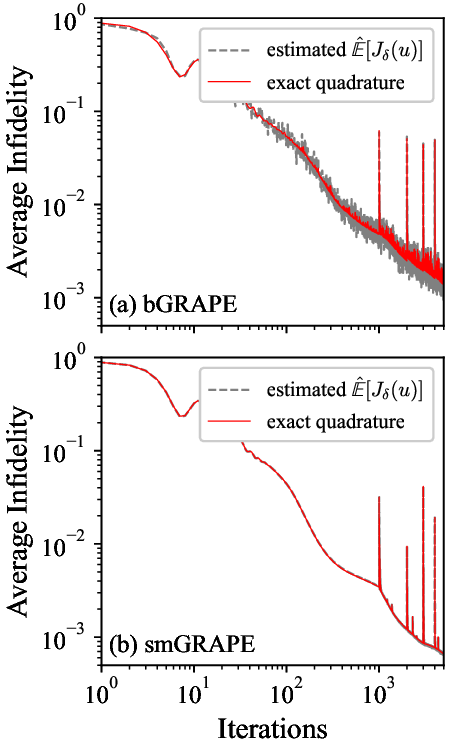}
     \caption{Comparison of the estimated average infidelity $\hat{\mathbb{E}}[J_\delta(u)]$ during the optimization between \textbf{bGRAPE} and \textbf{smGRAPE}. The exact quadrature is performed using high-order full dense grids using $5^5=3125$ points.}
    \label{fig:fig_estimated_error}
\end{figure}

We compare the control performance between the \textbf{smGRAPE} algorithm with $K=3$ and the \textbf{bGRAPE} algorithm \cite{wub-GRAPE2019} within 5000 iterations starting from the same initial conditions as shown in Figure~\ref{fig:fig_estimated_error}. The two algorithms intrinsically differ in their approach to approximating the average infidelity, as analyzed in Section \ref{sec:smROC}. 
It can be seen in Figure~\ref{fig:fig_estimated_error} that, when the size of the sampling set is determined to be $N=2d^2+2d+1=61$ for both algorithms, the estimated average infidelity $\hat{\mathbb{E}}[J_\delta(u)]$ obtained by the Smolyak sparse grid assisted method exhibits higher precision compared to that obtained by the Monte Carlo based method. An accurate estimation can effectively reduce the instability that may arise in subsequent optimization processes, thereby facilitating faster convergence. It should be noted that the observed abrupt increase in the optimization trajectory is mainly attributed to the characteristics of the Adam optimizer. Consequently, \textbf{smGRAPE} manages to provide the average infidelity of the value $6.0\times 10^{-4}$, while \textbf{bGRAPE} can only reach $1.1\times 10^{-3}$.

\subsection{Robust realization of the CNOT gate}

In this section, we focus on the robust realization of the CNOT gate (i.e. $U_{\rm{CNOT}}=\ketbra{00}+\ketbra{01}+\ketbra{11}{10}+\ketbra{10}{11}$), without correcting errors using local $z$ phases. The system Hamiltonian can then be given by
\begin{align} 
    H(t)=&\sigma_{z,1}\sigma_{z,2}+\Delta_1\sigma_{z,1}+\Delta_2 \sigma_{z,2}\nonumber\\
    &+\sum_{j=1}^2\left[u_{x,j}(t)\sigma_{x,j}+u_{y,j}(t)\sigma_{y,j}\right],
\end{align}
where $\sigma_{\{x,y,z\},j}$ are the Pauli operators corresponding to the $j$-th qubit, in the presence of system uncertainties described by $\Delta_1$ and $\Delta_2$ appearing on each qubit. Here, the uncertainties are assumed to follow the normal distribution, namely $\Delta_1,\Delta_2\sim{\mathbb N}(0,\varsigma^2)$. We consider the performance index defined in Eq.~\eqref{eq:infide_corr2} and apply piecewise control with $100$ segmentation over the time interval $T=4$.

\begin{figure*}
    \centering
    \includegraphics[width=0.9\textwidth]{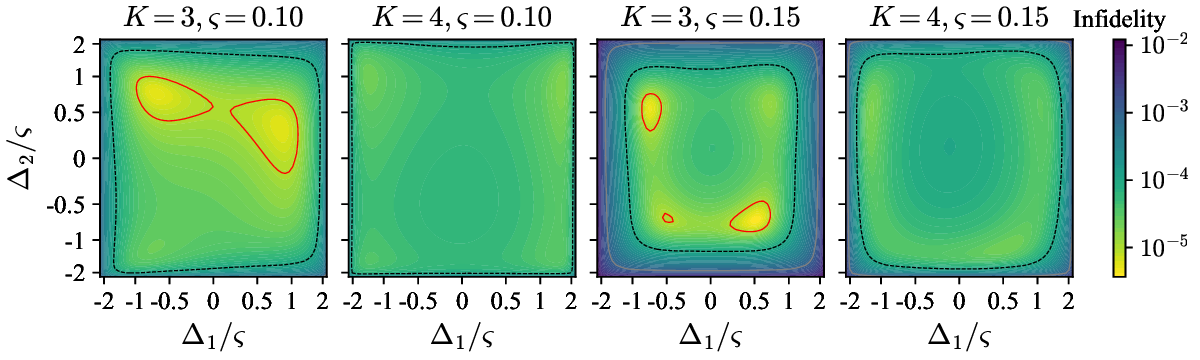}
    \caption{Landscapes demonstrating the control performance of the CNOT gate using \textbf{smGRAPE} against static detunings $\Delta_1$ and $\Delta_2$, with the sparse level $K=3,4$ and deviations $\varsigma=0.10,0.15$ (error correction with local $z$ phases is not involved). The gray solid lines, the black dashed lines and the red solid lines correspond to the infidelity values at $10^{-3}$, $10^{-4}$ and $10^{-5}$, respectively. The axes, corresponding to the values of $\varsigma$, are scaled according to the probability density of uncertainty in order to better interpret the robustness.}
    \label{fig:CNOT_robustness}
\end{figure*}

Since the Smolyak algorithm is essentially sample-based, choices of the sparse level $K$ and the parameterization $f(\veps)$ (as prescribed in Eq.~\eqref{eq:paraeps}) of $\delta_{1(2)}$ can influence the number and spatial position of samples. Experimentally for multiple uncertainties, it is not always true that the control performance of the Smolyak-assisted algorithm can be improved as the sparse level $K$ increases. For that reason, we compare the control performance of \textbf{smGRAPE} in different scenarios with $K=3,4$ and $\varsigma=0.10,0.15$, as demonstrated in Figure~\ref{fig:CNOT_robustness}. A higher value of the deviation $\varsigma$ which gives $\Delta_{1(2)}=\varsigma\veps$ ($\veps\sim \mathbb{N}(0,1)$) implies a wider range of samples, such that the region of infidelity bounded by $10^{-4}$ is larger (note that the true value along the axis is divided by $\varsigma$). When we keep the deviation value $\varsigma$, a higher value of the sparse level $K$ ($K=4$) also implies that more samples should be considered. Consequently, the average fidelity within a broader range can be improved, but minimum infidelity might increase.

\section{Conclusion}
\label{sec:conclusion}

In summary, inspired by Smolyak sparse grids, we develop a robust control scheme to deal with challenging difficulties in quantum technologies. This scheme can help improve the robustness of various quantum optimal control algorithms, including but not limited to \textbf{GRAPE} and \textbf{GOAT}. Compared to other robust quantum control schemes, the Smolyak algorithm assisted parametric scheme is promising to provide a universal sampling methodology to achieve strong robustness and high accuracy with less computational workload.

In more concrete terms, the Smolyak algorithm has been merged with gradient-based optimization methods such as \textbf{GOAT} and \textbf{GRAPE}, referred to as \textbf{smGOAT} and \textbf{smGRAPE} in our work, with successful applications to robust control problems related to quantum gates. For example, utilizing \textbf{smGOAT} or \textbf{smGRAPE}, the robustness of a single-qubit gate (the Hadamard gate, the $\pi/8$ gate, the phase gate or the $R_x(\pi)$ gate) or the CNOT gate can be greatly enhanced in the presence of uncertainties. During the optimization process, fewer samples are used and more stable convergence can be achieved compared to conventional methods such as \textbf{bGRAPE}.

Although the proposed sampling-based robust control approach presents superiority when handling low-to-medium uncertainty scales ($d\le10$), the number of samples is still large, especially when dealing with multiple complex uncertainties. And if the conditions of smoothness and bounded mixed derivatives are not satisfied, the performance of the Smolyak algorithm-assisted approach may degrade significantly. Our future work may include considering totally time-varying disturbances and further improving efficiency, contributing to the development of more reliable quantum computing and communication.

\begin{acknowledgments}
Z. Zhang and Z. Miao acknowledge support from the National Natural Science Foundation of China (Grant Nos. 62003113, 62173296, and 62273016), the Guangdong Basic and Applied Basic Research Foundation (Grant No. 2025A1515010186), and partial support from the Science Center Program of the National Natural Science Foundation of China under Grant 62188101. X.-H. Deng acknowledges support from the Shenzhen Science and Technology Program (Grant No. KQTD20200820113010023).
\end{acknowledgments}

\nocite{*}
\bibliography{apssamp}

\end{document}